\documentclass[prl,twocolumn,superscriptaddress,floatfix,noshowpacs]{revtex4}
\usepackage{amsmath,amsfonts,amssymb}
\usepackage{enumerate}
\usepackage{graphicx}
\usepackage{textcomp}
\usepackage{wasysym}
\usepackage{{appendix}}

\begin{document}

\title{Bursty egocentric network evolution in Skype}

\author{R. Kikas}
\affiliation{Software Technology and Applications Competence Centre (STACC), Estonia}
\author{M. Dumas}
\affiliation{Institute of Computer Science, University of Tartu, Estonia}
\author{M. Karsai \footnote{marton.karsai@aalto.fi}} 
\affiliation{BECS, School of Science, Aalto University, Finland}
\date{\today}

\begin{abstract}
In this study we analyze the dynamics of the contact list evolution of millions of users of the Skype communication network. We find that egocentric networks evolve heterogeneously in time as events of edge additions and deletions of individuals are grouped in long bursty clusters, which are separated by long inactive periods. We classify users by their link creation dynamics and show that bursty peaks of contact additions are likely to appear shortly after user account creation. We also study possible relations between bursty contact addition activity and other user-initiated actions like free and paid service adoption events. We show that bursts of contact additions are associated with increases in activity and adoption - an observation that can inform the design of targeted marketing tactics.
\end{abstract}

\maketitle

\section{Introduction}
\label{sec:intro}

The structure and evolution of human interactions are generally characterized by heterogeneity in manifold ways and are influenced by correlations ranging from individual level to global scale \cite{Barabasi2005a,Goh2008,Eckmann2004,Malmgren2008,Stehl2010}. Some of these emerging heterogeneities have been identified as the result of simple processes driven by microscopic rules. For example, preferential attachment \cite{Albert2002} has been shown to introduce degree heterogeneity and short path-ways into an evolving network structure. At the same time studies of action sequences -- such as communication actions in social networks -- have put into evidence further mechanisms responsible for inhomogeneities in network structure and dynamics \cite{Karsai2011a,Miritello2011,Kivela2012,Ko2012,Jin2012}. It has been shown in particular that correlated dynamics of individuals induce bursty temporal patterns of interactions \cite{Karsai2012a,Karsai2012b,Kovanen2011,Rybski2009}. However, for the most, these studies draw their conclusions from observation of static network snapshots or from incomplete temporal sequences of interactions. Only recently a number of datasets have been collected that contain time-stamped records of all topological actions, such as addition and deletion of edges between pairs of users \cite{Leskovec2008c,Backstrom2008,Gaito2012}. This development has opened the possibility to study directly the governing microscopic rules of network evolution in order to confirm previous hypothesis and to explore new mechanisms.

In particular, the availability of time-stamped records of user registrations, edge additions and edge deletions, allows to explore in details the microscopic evolution of \emph{egocentric networks} consisting of an individual user and their immediate friends or ``contacts''. Such datasets allow us to determine if the evolution of egocentric networks is incremental (small constant changes) or instead is characterized by sudden changes  (bursts). Sudden changes in the egocentric network could be induced by changes in the ego's social status (e.g. moving to a new place or starting a school) but also by the adoption of new services which open possibilities for alternative ways of interactions.

In this study we characterize the temporal evolution of egocentric networks in a very large online social network. We present for the first time empirical results on a dataset that contains anonymized data of hundreds of millions of subscribers of Skype -- one of the largest world-wide online communication system available. We mainly focus on the temporal evolution of social links of individuals and we detect correlated bursty periods in their edge addition and deletion dynamics. We also highlight some possible reasons behind bursty behaviour by looking for correlations between the observed dynamics and other user-initiated actions. To study these phenomena is not only important because we gain deeper understanding about technology-enabled human behaviour. It also provides insights to design marketing tactics, such as taking advantage of bursts in order to promote further user engagement (e.g. advertising new services when a user undergoes a burst in their egocentric network).

The rest of the paper is structured as follows. In Section \ref{sec:data} we give a brief description of the utilized datasets which after in Sections \ref{sec:egonet} and \ref{sec:corr} we present our main results about the temporal evolution and correlations of egocentric networks. In Section \ref{sec:rw} we overview the related works reported in the literature and finally in Section \ref{sec:concl} we summarize our main findings.

\section{Data}
\label{sec:data}

This research is based on a dataset consisting of a temporally detailed description of the social network of (anonymized) Skype users. 
For each user, the dataset provides the following details:
\begin{itemize}
\item Date of registration of the user
\item For each type of paid service (e.g. PSTN calls), date when the user first and last used this service (whenever applicable). 
\item Time series indicating the number of days in each month when the user connected to the Skype network
\item For each type of free service (e.g. Skype-to-Skype audio calls, video calls, chat, etc.), time series indicating the number of days in each month when the user used this service.
\item Time stamped events of link addition and deletion of each users.
\end{itemize}

In the Skype network, when a user adds a friend to his/her contact list, the friend may confirm the contact invitation or not. Also, at any point in time a user may delete a ``friend'' from their contact list.\footnote{There is also a possibility of ``blocking'' a user, but for the purposes of this research, contact blocking is treated as equivalent to contact deletion since the additional implications of blocking are irrelevant to the study.} Thus, the network evolves by means of the following events: contact addition, contact confirmation and contact deletion. In our study to take into account only trusted social links we retained only confirmed edges, meaning edges where both parties accepted the connection. Failure to do so would lead to mixing undesired with desired connections. 

For the present study we employed two subsets of the above dataset. The first dataset (DS1) includes every active user as of the end of 2010, all confirmed edges between these users, and the date of confirmation of each edge. In this context, we define an active user as one who connected to the Skype network at least in two different months during the first year after their registration date. In order to consider users with realistic number of friends we selected only those with degree between $2$ and $1000$. Users with more connections are suspected to be bots or are business accounts and their behavior differs from the majority who use Skype for personal communication. This filtering led to a set of more than 150 million users.

The second dataset (DS2) includes the set of edge addition events and edge confirmation events for the period of $2009-2011$ as well as edge deletion events recorded for a year-long period in $2010-2011$. Each of these events is time-stamped with a date. Only events related to users with degree between $2$ and $1000$ were kept. Unlike DS1, non-active users were retained in DS2.

\section{Egocentric network evolution}
\label{sec:egonet}
In this section we look at the evolution of egocentric networks to gain deeper understanding about the governing microscopic rules of contact list evolution.

\subsection{Bursty edge dynamics}
\label{sec:bursty}

To characterize the temporal evolution of contact lists, we first examine the sequences of edge addition and deletion of each individual and we calculate the distributions of inter-event times
\begin{equation}
\tau_{a}=t^{a}_{i+1}-t^{a}_{i} \hspace{.2in} \mbox{,} \hspace{.2in} \tau_{d}=t^{d}_{i+1}-t^{d}_{i} \hspace{.2in} \mbox{,} \hspace{.2in} \tau_{ad}=t^{d}_{e}-t^{a}_{e}
\label{eq:1}
\end{equation}
elapsed between consecutive additions at $t^{a}_{i}$ and $t^{a}_{i+1}$ or deletions at $t^{d}_{i}$ and $t^{d}_{i+1}$ of the same user or between the addition and deletion of an edge $e$. If this distribution follows a power-law as
\begin{equation}
P(\tau)\sim \tau^{-\gamma}
\end{equation}
it indicates strong temporal heterogeneities and burstiness, or otherwise if it decays exponentially it reflects regular dynamical features. Bursty temporal evolution of human dynamics was confirmed in various systems ranging from library loans to human communication \cite{Goh2008,Karsai2012b} or recently for the evolution of social networks \cite{Gaito2012}.

\begin{figure}[ht!]
  \centering
  \includegraphics[width=7cm]{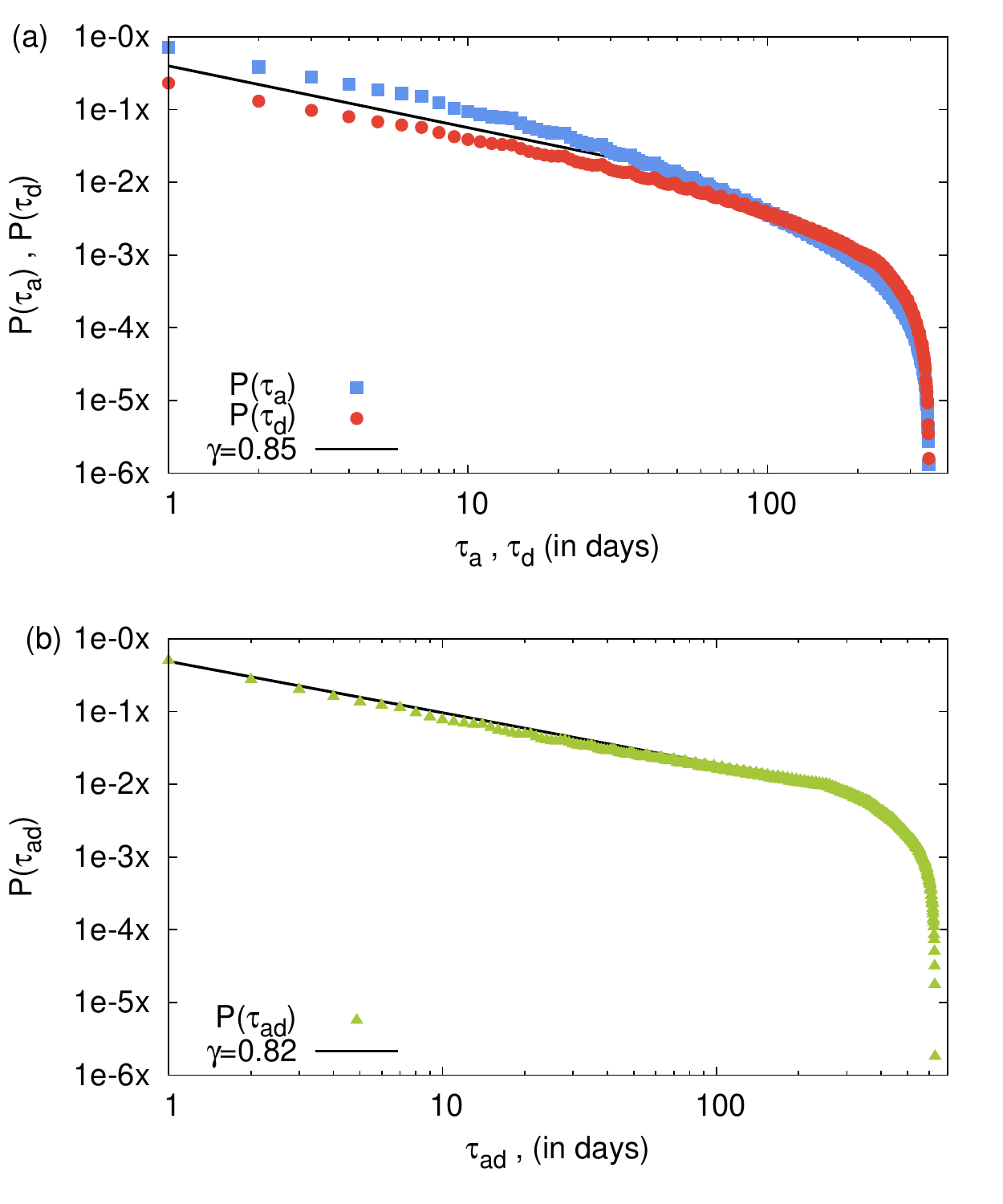}
  \caption{Inter-event time distributions of (a) edge addition (blue squares) and deletion (red circles) events of users and (b) addition and deletion of links (green triangles) in DS2. The straight lines indicate power-law functions with exponent (a) $\gamma=0.85$ and (b) $\gamma=0.82$. For a formal definition of the $\tau_a$, $\tau_d$, and $\tau_{ad}$ see Eq.\ref{eq:1}.}
 \label{fig:deletion_interevent_pdf}
\end{figure}

Calculating $P(\tau)$ for DS2 we observe heterogeneously distributed inter-event times in Fig.\ref{fig:deletion_interevent_pdf}.a both in case of edge additions and edge deletions. The distributions are showing rather similar scaling with a section fitting on a power-law with exponent $\gamma\simeq 0.85$ and an exponential cutoff due to the finite time window. This is an interesting observation as one would expect rather different decision mechanisms behind adding and deleting a contact. Specifically, one would expect that contact addition is driven by the desire or need to communicate or to signal a social relation, while edge deletion is driven by the desire not to be visible or accessible by the deleted contact. Moreover, the edge life-time distribution (in Fig.\ref{fig:deletion_interevent_pdf}.b) can be characterized by a power-law function also with exponent $\gamma=0.82$ and an exponential cutoff due to finite-size effects.

The similarity of these distributions indicates common temporal features and they provide evidence of bursty dynamics both in case of edge addition and deletion as well as for edge life-times. Below we examine this bursty dynamics more in details.

\subsection{Trains of bursts}

The inter-event time distributions in Fig.\ref{fig:deletion_interevent_pdf}.a are indicative of the presence of temporal heterogeneities, however they cannot show whether further correlations are present between consecutive actions. Recently a new methodology was developed \cite{Karsai2012a} to address this question and evince evident correlations in heterogeneous binary signals. The crux of the method is to group consecutive events, which follow each other with inter-event times smaller or equal to $\Delta t$, into bursty event cluster (trains). The sensitive measure of correlations serves as the $E$ number of events in  bursty clusters. Their distribution scales as a power-law
\begin{equation}
P(E)\sim E^{-\beta}
\end{equation}
if the signal is correlated and long event trains are evolving in the dynamics. On the other hand if consecutive events are independent it decays exponentially even when the inter-event time distribution is fat-tailed.

\begin{figure}[ht!]
\centering
\includegraphics[width=7cm]{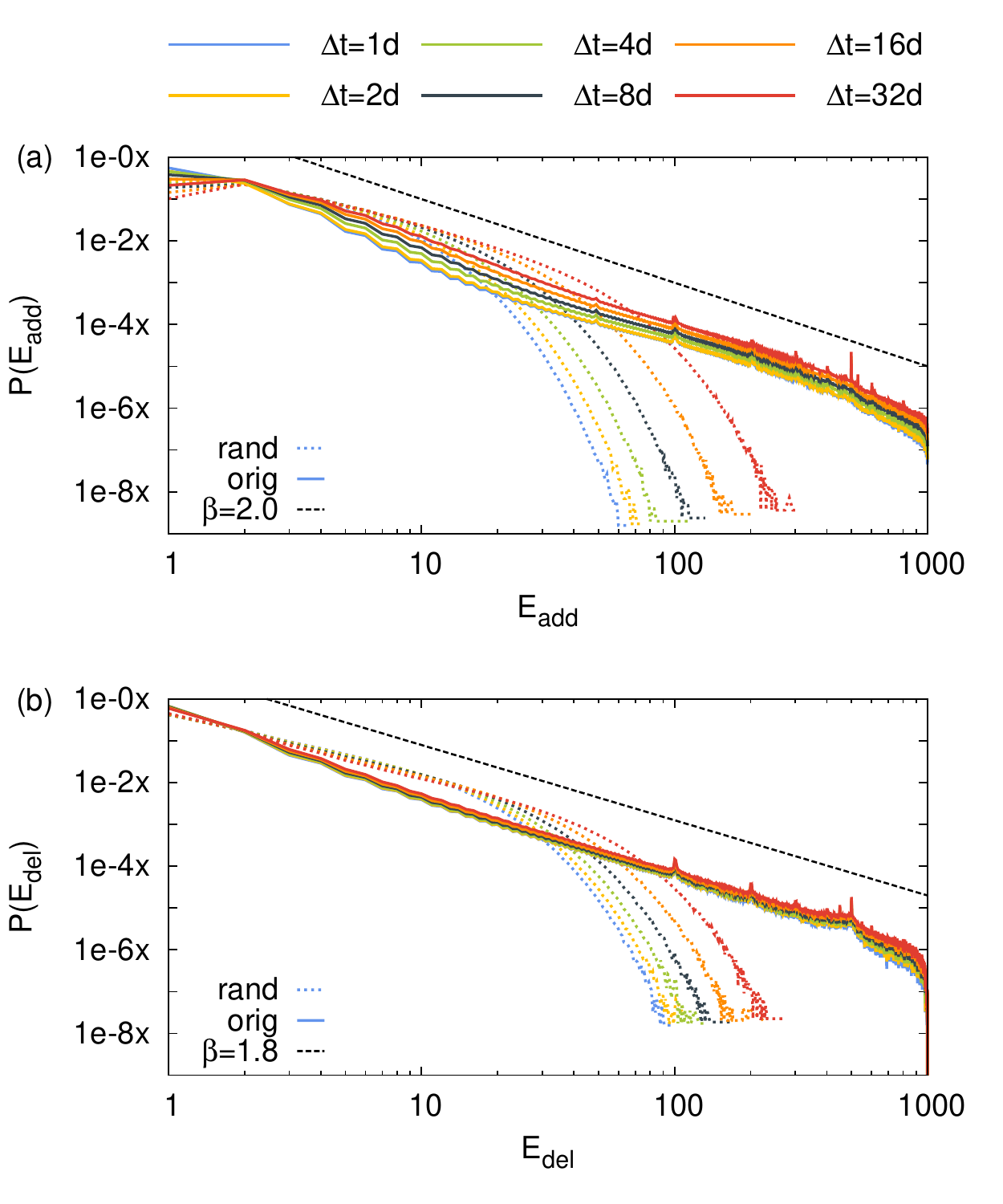}
\caption{Distribution of number of events in bursty trains of (a) contact addition and (b) deletion of individuals in DS2. Distributions were calculated with time window sizes $\Delta t=1, 2, 4, 8, 16$ and $32$ days. Distributions calculated for randomly shuffled sequences are also presented (dashed line) calculated with the same $\Delta t$ values. Straight lines are indication of power-law functions with exponents (a) $\beta=2.0$ and (b) $\beta=1.8$.}
 \label{fig:PE}
\end{figure}

In the present case we analyzed the edge modification sequence of each individual by extracting the clusters of events of new edge addition and deletion (the trains) and recording their size $E_a$ ($E_d$). The fact that the $P(E_a)$ and $P(E_d)$ distributions in Fig.\ref{fig:PE} span over orders of magnitude confirm the presence of correlations evolving between consecutive events of edge additions (deletions). The corresponding train sizes are distributed as a power-law with characteristic exponent values $\beta_a\simeq 2.0$ ($\beta_d\simeq 1.8$). This scaling behaviour appears to be robust against the choice of the $\Delta t$ window size as it remains similar for distributions calculated with $\Delta t=1, 2, 4, 8, 16$ and $32$ days.

The presence of correlations in the egocentric dynamics are even more apparent if we compare the empirical $P(E)$ functions to the equivalent distributions calculated for independent signals. To receive a reference system like this we used the $\tau_a$ ($\tau_d$) inter-event times of the original sequences, put them in a pool and redraw for each user randomly as many inter-event times as they had originally. The inter-event time distribution of the resulting randomly-shuffled null model was the same as the original $P(\tau)$ distribution as we used the same $\tau_a$ ($\tau_d$) values and also the shuffled sequences of each user contained the same number of events as before. However, with this random shuffling method we destroyed all possible temporal correlations which were present between the consecutive events of single users. The $P(E)$ distribution of such randomly-shuffled sequences should decay exponentially \cite{Karsai2012a} and any discrepancy from this scaling behaviour is indicative of correlations. 

It is demonstrated in Fig.\ref{fig:PE}.a and b that the $P(E)$ distributions calculated for the randomly-shuffled reference sequences (dash\-ed lines) are exponentially distributed and they are very different from the original distributions (solid lines). It puts into evidence that the actions of an individual are not independent and we can conclude that the evolution of egocentric networks is not only heterogeneous in time but also that intrinsic correlations are driving its dynamics. They lead to the presence of high activity bursty periods, where a large number of edges are added or deleted, and which are followed by long low activity intervals.

\section{Groups of individual dynamics}
\label{sec:corr}

\begin{figure*}[ht!]
\centering
\includegraphics[width=14cm]{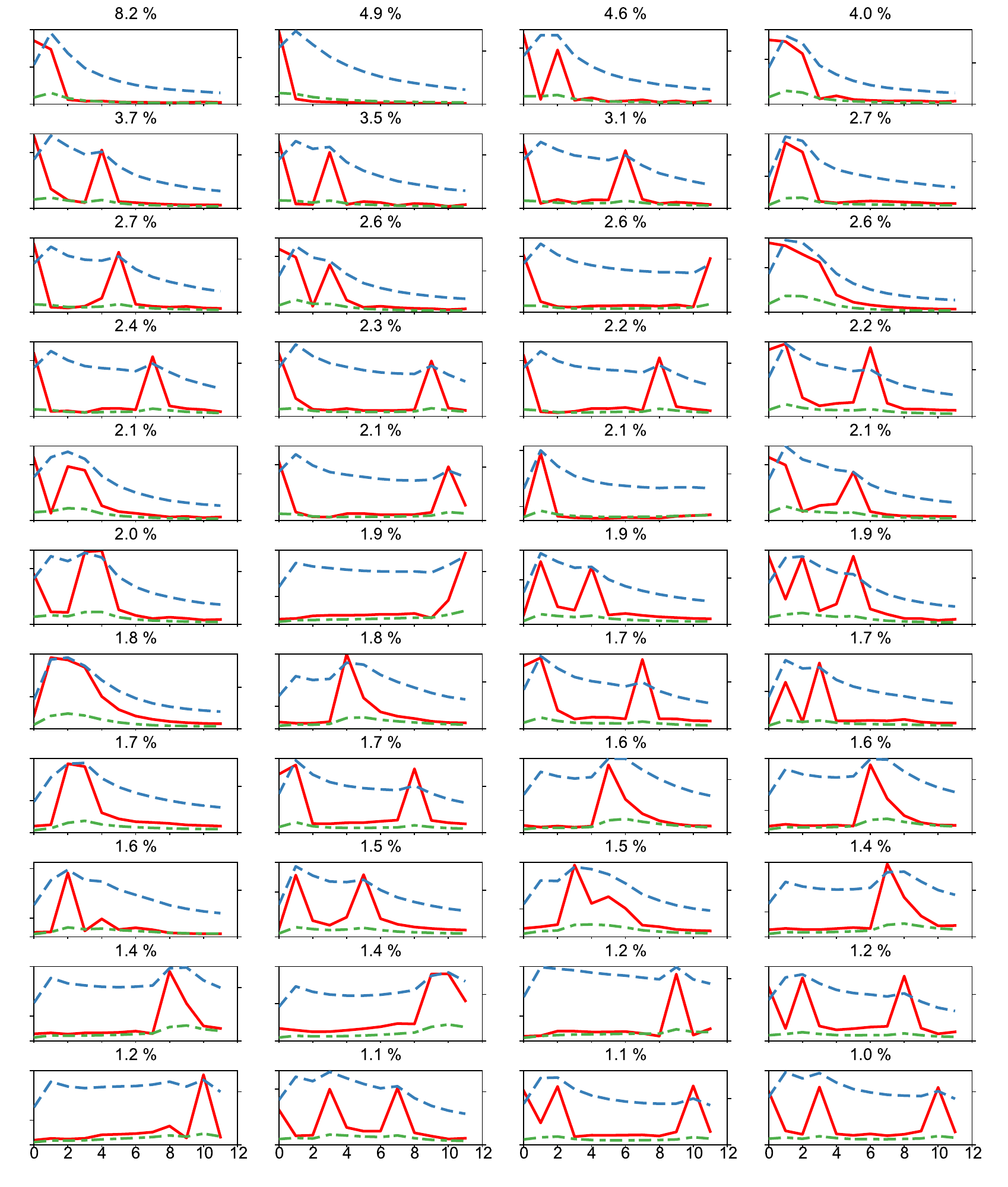}
\caption{Characteristic groups of contact addition patterns calculated for DS1 by using the SAX and k-means clustering methods. On each panel red solid lines correspond to the average number of new contacts added at the actual month, dashed blue lines are the average number of connected days, while green pointed lines are the average number of days used Skype-to-Skype communication belonging to the actual group. Left scales corresponds to the number of added contacts while right scales are number of days. The ticks on the left (right) vertical scale is corresponding to the same value on each panel.}
\label{fig:patts}
\end{figure*}

So far we have observed that edge addition and deletion events of an individual are bursty and clustered in time, yet we know less about when these bursty trains are evolving during the lifetime of a user. Do they appear in any time or there are typical activity patterns of edge additions or maybe triggered by other user actions? In the following we address these questions by seeking correlations of bursty peaks with other user activities.

\subsection{Grouping methods}

In order to compare the edge addition sequences of individuals we used DS1 and concentrated only on the activity of users during the first year of their $t_u$ user time i.e. the time after their registration. We were keeping track the $a_i$ number of newly added edges of each node $i$ with a single month resolution and receive a discrete $a_i(t_u)$ sequence for each individual where $t_u=1...12$. 

Note that using a granularity of one day for this part of the study  leads to activity curves with strong fluctuations, which could be explained by factors that affect individual behaviour at the fine temporal level. Such effects are attributable to daily and weekly fluctuations in human activity, public holidays, service outages, etc. These fluctuations are sufficiently heavy and manifold to prevent us from detecting meaningful correlations and clear groupings of typical user behaviour with granularity lower than the month.

To be able to compare sequences of users with diverse overall intensity we applied the \textit{Symbolic Aggregate Approximation} (SAX) method \cite{Lin2003} with alphabet size 10. This method was selected because it keeps the characteristic shape of the activity function of each individual but it makes them comparable via a normalization method. 
The method starts by normalizing each sequence independently  to have a zero mean and unit variance. The normalized time series follows Gaussian distribution  \cite{Lin2003}  and using this fact, we can discretize the time series values by finding breakpoints that will produce 10 (\emph{the alphabet size}) equal-sized areas under the Gaussian $N(0,1)$ curve. Original values are replaced based on the breakpoints: all values lower than the smallest breakpoint will receive 0, values between the first  and second breakpoint will receive 1, etc. The resulting sequence has only 10 different values. The discretization is meant to maintain the shape and characteristic peaks of the sequence while eliminating differences between value ranges.

To detect groups of users with similar edge addition dynamics after performing normalization via SAX, we applied the \textit{k-means clustering}  method \cite{Hartigan1979} on the activity sequences using euclidean distance. K-means algorithm was chosen for its ability to scale to $O(100M)$ data points. Also, it has been shown that running k-means on data that was previously processed by SAX produces better results than the original data \cite{Lin2007}. To choose the optimal number of clusters we executed the clustering method for different $k$ values up to 200 and we determined that the sum of squared errors levels off at $k>40$ (this approach is also referred as the \textit{Elbow method}). After inspecting different clustering configurations by using several initial random seeds, we determined $k=44$ to be around optimal.

\subsection{Empirical results}

\begin{figure}[hb!]
\centering
\includegraphics[width=7cm]{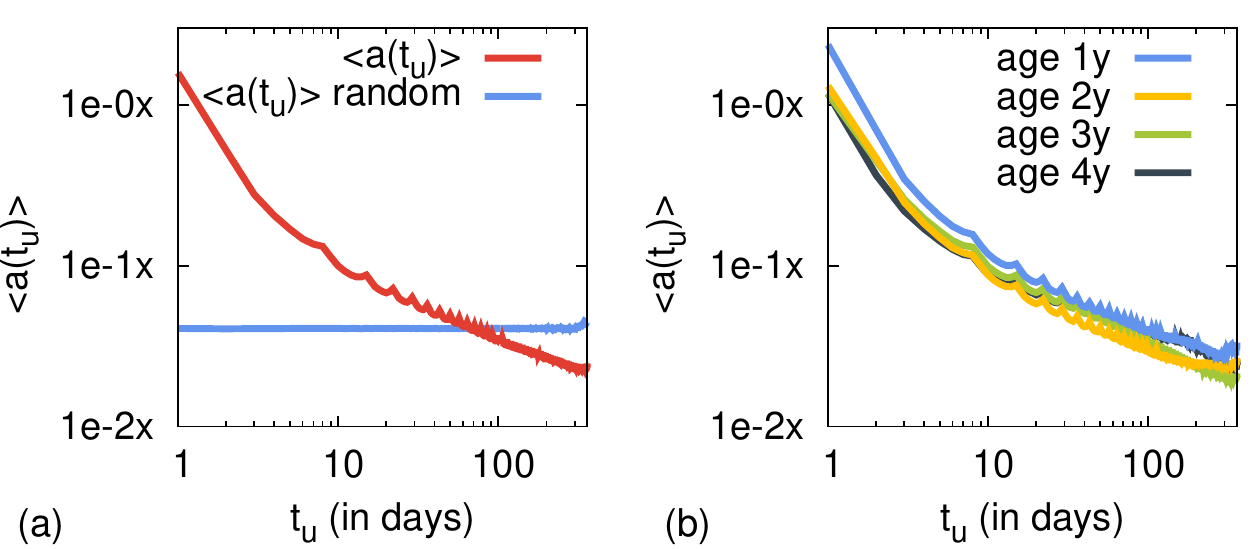}
\caption{Average number of new contacts added as function of $t_u$ user time calculated for (a) all users in DS1 and (b) for users grouped by the time they joined Skype. Age in the (b) figure caption refers to the time of registration in advance of the measurement.}
\label{fig:newdegevage}
\end{figure}

In Fig.\ref{fig:patts} we show the $\langle a_i(t_u) \rangle_k$ average activity curves of each $k$ cluster together with the percentage of users who are belonging to the actual group (panels are in descending orders with respect to the size of the cluster of users they cover). Clustering is obtained by considering only contact addition (red line). Looking at the most common patterns it is straightforward that typically people perform their principal (the largest and usually the only one) edge addition burst right after they join the network. This is the time when they explore their social acquaintances who have already joined Skype before and which after they add contacts just occasionally with lower frequency. This behaviour is confirmed by looking at the $\langle a_i(t_u) \rangle$ average number of new edge additions (Fig.\ref{fig:newdegevage}.a) calculated for each user. Note that a similar behaviour was observed in other studies \cite{Gaito2012}. In addition, Fig.\ref{fig:newdegevage}.b shows that this correlation is independent from the time of registration as similar early stage peaks were found for users who joined Skype at different years.

To verify that this observation is not tied to the chosen method for preprocessing the activity functions, we experimented with a second method that combines standardization with \textit{Discrete Wavelet Transform} \cite{Chaovalit2011}. The clusters obtained with this alternative method were different, but the above phenomenon can still be observed (see Appendix).

To check the significance of this phenomenon we compare the $\langle a_i(t_u) \rangle$ overall average curve to a similar curve $\langle a_i(t_u) \rangle_r$ calculated for independent sequences. To generate the null model sequences we apply a very similar method as earlier. We take the activity values of each user at each month, randomly shuffle them and redistribute between users. This way the overall activity remains unchanged and each user have a sequence of $12$ data points as before, but the correlations between activity peaks and user time are destroyed. The resulting average activity curve becomes flat evidently as it is demonstrated in Fig.\ref{fig:newdegevage}.a (blue line). Comparing the original and random curves (red and blue lines in Fig.\ref{fig:newdegevage}.a) it is straightforward that the peak at early times, which is visible for the original curve, is not observable for the null model curve, thus supporting the significance of the correlation.

\section{Correlations at later times}

A single bursty peak at early time is not the characteristic of every user. Less common motifs in Fig.\ref{fig:patts} show that principal bursts may emerge later or even in multiple times. This observation indicates that events other than user registration may also trigger immediate changes in the egocentric network. Changes in social status as moving to another place or starting school could be possible reasons behind the later bursty peaks, but given the actual dataset, we are not in a position to confirm these effects. However, it is possible to study dependencies between the dynamics of contact addition and other system-related activities.

\begin{table}
\begin{center}
\begin{tabular}{|c|c|c|c|c|}
\hline  & S2S & activ & S2S rand & activ rand \\ 
\hline $\langle r \rangle$ & $0.34608$ & $0.308137$ & $8.31909$e{-6} & $1.69978$e{-5} \\ 
\hline 
\end{tabular}
\end{center}
\caption{Average correlation coefficient calculated between edge addition dynamics-free service usage (S2S) and connected days (activ). Values received for random sequences are also presented.}
\label{table:1}
\end{table}

\begin{figure}[ht!]
  \centering
  \includegraphics[width=7cm]{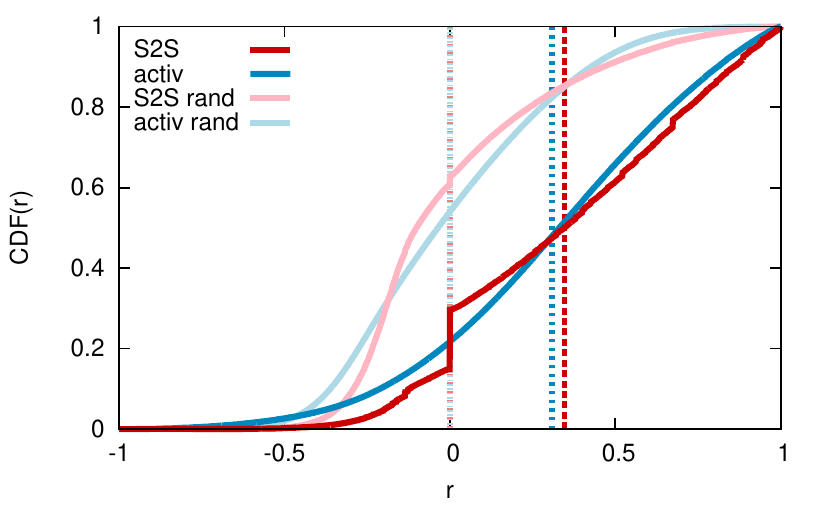}
  \caption{Cumulative distribution of the $r$ Pearson correlation coefficients calculated between the contact addition dynamics of individuals, their service usage intensity and user activity during their first $t_u=12$ months (solid dark red and blue lines accordingly). Similar curves calculated for random sequences are also shown (solid light red and blue accordingly). Dashed vertical lines assigns the average correlation coefficient $\langle r \rangle$ of the present distributions (for numerical values see Table.\ref{table:1}).}
 \label{fig:correlation_s2s}
\end{figure}

In Fig.\ref{fig:patts} besides the average edge addition rates we also show the average number of connected days and average number of days of free-service usage for each group. Comparing these curves one can foresee some dependency between them as users are performing bursty contact addition months at the time when they are connected and also heavily using free services. To quantify these relationships we calculated a correlation coefficients for each user $i$ defined as
\begin{equation}
r^s_i=\frac{\langle (a_i(t)-\bar{a}_i)(s_i(t)-\bar{s}_i) \rangle_t}{\sigma_{a_i} \sigma_{s_i}}
\end{equation}
where $s_i(t)$ denotes the sequence of number of connection days or free service usage days and the average is running through $12$ discrete time steps. The $CDF(r)$ cumulative distribution of the two coefficients calculated for every users is depicted in Fig.\ref{fig:correlation_s2s} (dark red curve for correlations with Skype-to-Skype free services (S2S) and dark blue curve with connected days (activ)). They indicate mostly positive correlations as almost no user was found with coefficient $r\leq -0.5$ and at the same time approximately $80\%$ of users present non-negative correlations in both cases. The $\langle r \rangle$ average correlation coefficients of the two distributions assign also strong positive correlations as they take values $\langle r_{S2S} \rangle = 0.34608$ and $\langle r_{activ} \rangle = 0.308137$ for the free service and user activity accordingly (vertical dashed lines in Fig.\ref{fig:correlation_s2s}). 

To check whether the observed positive correlations are significant or only the results of random fluctuations of independent processes we calculated the correlations between the same curves but after we randomly shuffled the activity sequences. To generate the null model sequences we applied the same method described above in Section \ref{sec:corr} for each sequences. The cumulative distribution functions of the correlation coefficients calculated between the random sequences are shown in Fig.\ref{fig:correlation_s2s} (light red line for free service usage and light blue line for user activity). These curves are very different from the ones of original sequences and also their average values are presenting significant discrepancies. For random sequences $\langle r_{S2S}^{rand} \rangle \simeq \langle r_{activ}^{rand} \rangle \simeq 0$ in agreement with the expected values of correlations calculated between independent signals (the average values for all calculated $CDF(r)$ are summarized in Table \ref{table:1}). Consequently the observed dependencies between the original sequences are significant and they indicate true positive correlations between the edge addition dynamics of individuals, their user activity and free service usage.

\begin{figure}[ht!]
  \centering
  \includegraphics[width=7cm]{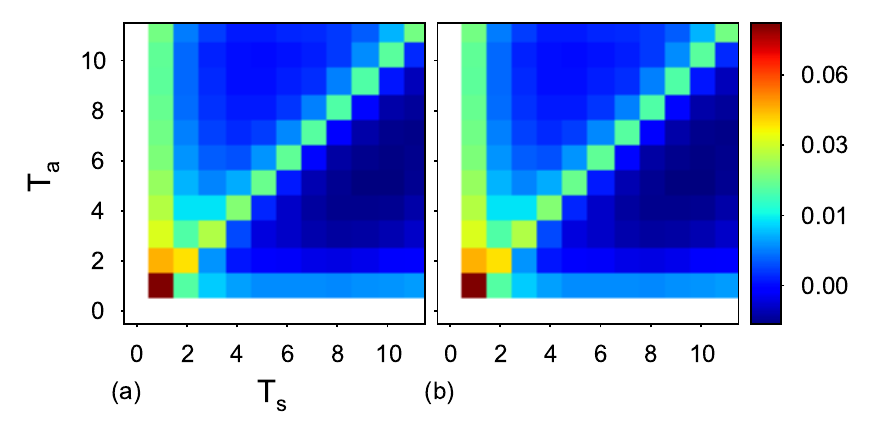}
  \caption{Matrices of conditional probabilities that a user performed a peak contact addition month at $T_a$ if he adopted a (a) free or a (b) paid service at $T_s$. Probability values at $T_a=0$ and $T_s=0$ are not shown. Colors are coding logarithmically the probability values.}
 \label{fig:correlation_payg_adoption}
\end{figure}

Another possible reason behind sudden changes in the egocentric graph can be due to the adoption of a new communication service which opens an alternative way of interaction. This channel could be a free communication service what the user explored or could be a paid service what he/she subscribed for. Therefore, one can look for the time when a user starts to use a free or paid service for the first time and check whether these actions can trigger bursty peaks in contact addition. To do so we identify bursty peak months for each user as the months where the contact addition activity is
\begin{equation}
a^p_i(t)=(a_i(t)|a_i(t)\geq \bar{a}_i+2\sigma_{a_i}).
\end{equation}
Here $\bar{a}_i$ and $\sigma_{a_i}$ denotes the average and standard deviation of the contact addition activity sequence of user $i$. In Fig.\ref{fig:correlation_payg_adoption}.a and b we present the conditional probabilities that a user performs a peak month of contact addition at a given user time $T_a$ if he/she adopted a free (Fig.\ref{fig:correlation_payg_adoption}.a) or a paid service (Fig.\ref{fig:correlation_payg_adoption}.b) at time $T_s$. We have seen earlier that strong correlations are playing role between the user activity and registration time at $t_u=0$. However, here we are looking for correlations which evolve later in user time. To avoid the dominating effect of strongly correlated early activity peaks in Fig.\ref{fig:correlation_payg_adoption} we neglect the data bins belonging to adoption time $T_s=0$ and bursty peak time $T_a=0$ and normalize the probabilities accordingly. After this preparation, correlations between late service adoption and peak contact addition months became visible as a high activity diagonal appeared in the probability matrices in Fig.\ref{fig:correlation_payg_adoption}.a and b. Consequently together with correlations with registration and user activity this result serves us another possible explanation for the late evolution of bursty contact addition peaks as they are possibly triggered by paid or free service adoption.

\section{Related work}
\label{sec:rw}

Temporal evolution of networks was studied thoroughly during the last years as datasets recording the dynamics of millions of interacting entities became available \cite{Holme2012}. One of the most investigated area was the evolution of large social networks \cite{Krings2009,Krings2012,Onnela2007,Goncalves2011} where it has been shown that several mechanisms push such networks towards developing heterogeneous topologies and strongly modular structures \cite{Albert2002,Leskovec2008c,Backstrom2008}. In addition various methodologies have been developed to detect evolving mezoscopic patterns \cite{Milo2002,Kovanen2011} and emerging community structures \cite{Fortunato2010}. Our study falls under the same umbrella as these previous works but focuses on the temporal evolution of egocentric networks.

Heterogeneities in the dynamics of social interactions have been observed by following the communication sequences of individuals \cite{Barabasi2005a,Goh2008,Eckmann2004,Malmgren2008,Stehl2010}. Circadian fluctuations and long range temporal correlations were shown to play important role here \cite{Jo2012,Karsai2012a,Karsai2012b,Rybski2009} and they partially explain the observed non-homogeneous behaviour. Lately heterogeneous evolution of social networks was also reported by Gaito et.al. \cite{Gaito2012} who analyzed the dynamics of the Renren online social network. In this paper \cite{Gaito2012} the authors simultaneously arrived to similar conclusions like us regarding the burstiness in the evolution of contact addition of users. In our study -- beyond confirming this effect in an independent dataset -- we extend this finding in two ways. First, we put forward evolving bursty trains also in the sequence of contact deletion of individuals and second we highlight non-trivial correlations triggering bursty periods in the evolution of egocentric networks.

\section{Conclusions}
\label{sec:concl}

We investigated the temporal evolution of egocentric networks in one of the largest online social networks available, namely the Skype network. Our main observation was that the dynamics of edge addition and deletion show strongly heterogeneous temporal behaviour as most of the edges are added or deleted during very short bursty periods, which are separated by long low activity intervals. During such high activity periods long bursty trains of contact addition events can evolve confirming the presence of intrinsic correlations. We also concluded that such trains show strong relation with the registration time as they are most likely to appear right after the user joined the network. High activity bursty peaks, which evolved later and even in multiple times were also detected for some users.

We showed that such patterns are correlated with user activity and free service usage and could be triggered by free and paid service adoption. The observed temporal behaviour and non-trivial correlations disclose characteristics about the evolution of social networks, which suit well into the general picture of human dynamics as correlations and heterogeneity were confirmed earlier in many independent cases. However, beyond sophisticating our present assumptions about human behaviour, these results serve a more pragmatic advantage as they may help to improve the design of online services and marketing tactics, maybe used for more effective targeted advertisements.

{\bf Acknowledgements} The authors gratefully acknowledge the support of Andr\'{e} Karpi\v{s}t\v{s}enko and Ando Saabas from Skype Technologies. This research was partly funded by the ERDF via the Software Technology and Applications Competence Centre (STACC) and Skype Technologies. MK thanks K. Kaski and J. Kert\'esz for useful discussions and acknowledges support from FP7 ICTeCollective Project (No. 238597).

\begin{appendices}
\section{Appendix}
\subsection{Clustering by Discrete Wavelet Transform}

In the main text we applied the \textit{Symbolic Aggregate Approximation} (SAX) method to normalize the edge addition activity functions of individuals and categorised them by k-means clustering. However, we can apply other methods, which could provide different user clusters and may suggest different characteristic link addition dynamics. To investigate this option we repeated our measurements by calculating the \textit{Discrete Wavelet Transforms} (DWT) of individual curves and cluster users by k-means clustering.

\begin{figure}[htb]
\centering
\includegraphics[width=3.3in]{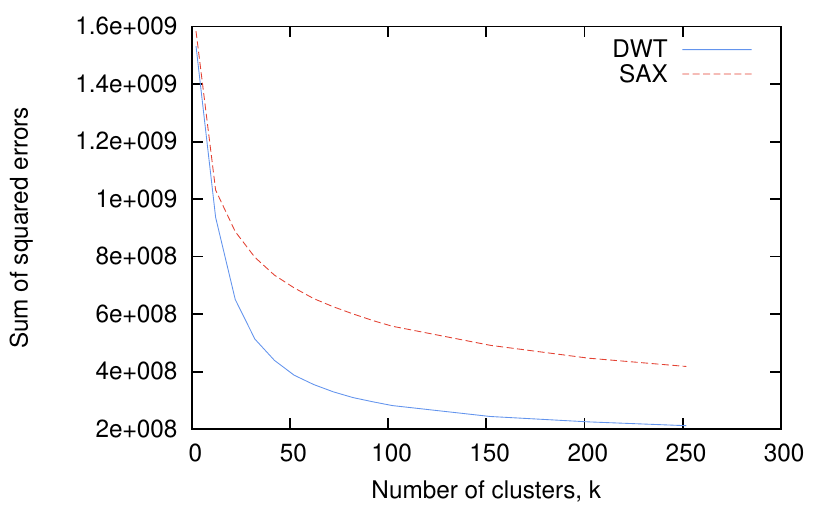}
\caption{Change of the sum of squared errors as a function of $k$ cluster number calculated for individual activity curves processed by SAX and DWT methods.}
\label{SMfig:1}
\end{figure}

We apply Discrete Wavelet Transform with the Haar wavelet \cite{Burrus} on the individual's activity vectors and obtain 16 coefficients. (Note that the Haar wavelet works only on sequences of length power of two, therefore we extended the sequence by padding with zeros from 12 to 16). For dimensionality reduction, we only use the first 8 coefficients. By removing the last 8 coefficients, we maintain the high level shape of the sequence. We calculate the groups by performing k-means clustering directly on the coefficients and use the \textit{Elbow method} to determine the optimal number of clusters. In Fig.\ref{SMfig:1} we demonstrate that even the sum of the squared errors of the calculations obtained by the SAX and DWT methods are very different. The optimal number of clusters (provided by the Elbow method) is close to be $40$ in both cases.

\begin{figure*}[ht!]
\centering
\includegraphics[width=14cm]{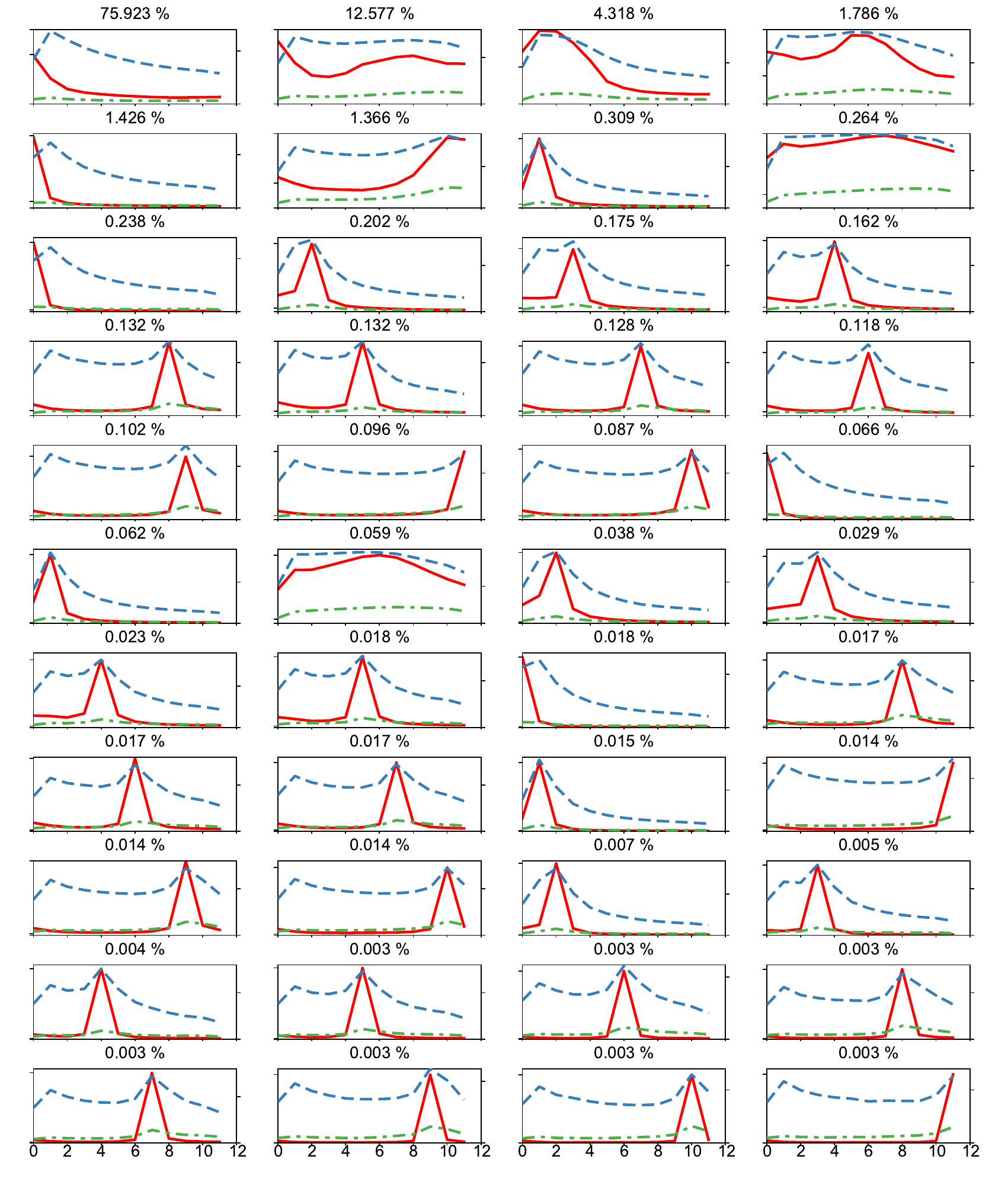}
\caption{Characteristic groups of contact addition patterns calculated for DS1 by using DWT and k-means clustering methods. On each panel red solid lines correspond to the average number of new contacts added at the actual month, dashed blue lines are the average number of connected days, while green pointed lines are the average number of days used Skype-to-Skype communication belonging to the actual group. Left scales corresponds to the number of added contacts while right scales are number of days. The ticks on the left (right) vertical scale is corresponding to the same value on each panel.}
\label{SMfig:2}
\end{figure*}

Finally by calculating the average edge addition curves for each clusters obtained by DWT and k-means clustering (see results in Fig.\ref{SMfig:2}) we can take the same conclusions as for the corresponding calculations applying the SAX method (in Fig.3 in the main text):
\begin{itemize}
\item the cluster(s) covering the highest proportion of users are those clusters that have a peak at the beginning
\item the peaks in contact addition are accompanied by increases in service usage.
\end{itemize}

\end{appendices}

\end{document}